\documentclass{ptapap}
\usepackage{color}

\usepackage{soul}

\usepackage{amsmath}
\usepackage{amssymb}

\usepackage[colorlinks=true]{hyperref}
\hypersetup{linkcolor=blue,citecolor=blue,filecolor=blue,urlcolor=blue}

\author{Amit Kumar Mandal}[CFT]
\author{Bożena Czerny}[CFT]
\author{Francisco Pozo Nuñez}[Heidelberg]
\author{Vikram Kumar Jaiswal}[CFT]
\author{Mohammad Hassan Naddaf}[Liege]
\author{Swayamtrupta Panda}[Gemini]
\author{Paulina Karczmarek}[CAMK]
\author{Grzegorz Pietrzyński}[CAMK]

\affil[CFT]{Center for Theoretical Physics, Polish Academy of Sciences, Al. Lotników 32/46, 02-668 Warsaw, Poland}
\affil[Varanasi]{Department of Physics, Institute of Science, Banaras Hindu University, Varanasi 221005, India}
\affil[Utah]{Department of Physics and Astronomy, University of Utah, Salt Lake City, UT 84112, USA}
\affil[Liege]{Institut d’Astrophysique et de Géophysique, Université de Liège, Allée du six Août 19c, B-4000 Liège (Sart-Tilman), Belgium}
\affil[Gemini]{International Gemini Observatory/NSF NOIRLab, Casilla 603, La Serena, Chile}
\affil[Heidelberg]{Astroinformatics, Heidelberg Institute for Theoretical Studies, Schloss-Wolfsbrunnenweg 35, 69118 Heidelberg, Germany}
\affil[CAMK]{Nicolaus Copernicus Astronomical Center, Polish Academy of Sciences, Bartycka 18, 00-716 Warszawa, Poland}

\title{HALO: An AGN Monitoring Program to Constrain the Hubble Constant}

\begin{document}

\maketitle

\begin{abstract}

We present a model that couples a standard accretion disk, an optional inner hot flow, lamp-post irradiation, a radiation-pressure–regulated BLR, and emission from a dusty torus. Parameterized by black hole mass, accretion rate, inclination, lamp-post height, and cloud properties, the model reproduces the spectral energy distribution (SED), the H$\beta$ BLR lag, and inter-band continuum delays. Simultaneous SED and lag-spectrum fitting with free accretion rate and luminosity distance, enables an estimate of the Hubble constant ($H_{0}$), demonstrating that inter-band delays offer strong cosmological leverage. To advance this approach, we have initiated a long-term AGN monitoring program with the 60 cm telescope at the Rolf Chini Cerro Murphy Observatory (OCM) in Chile. 

\end{abstract}

\section{Introduction}

The inflow of matter onto a supermassive black hole (SMBH) at the center of an active galactic nucleus (AGN) leads to the formation of an accretion disk. In standard thin-disk and lamp-post frameworks \citep{SS1973, 1989ESASP.296..945G, Matt1991}, a compact X-ray source irradiates the disk, whose reprocessed UV–optical emission arises from progressively larger radii as the temperature declines outward following $T_{\rm disk} \sim R_{\rm disk}^{-3/4}$.

Thus, AGN variability offers a direct probe of this structure. Reverberation mapping (RM) measures the time-delayed response of longer-wavelength emission to short-wavelength fluctuations, providing estimates of the continuum-emitting region \citep[e.g., continuum--RM;][]{cackett_2021iSci, edelson2024, 2025ApJ...985...30M, 2025A&A...700L...8P, 2026A&A...706A.176M, 2026A&A...714A.229P}. However, observed inter-band delays often exceed the predictions of power-law standard disk model \citep{fausnaugh2016, cackett2018}. A key cause is contamination of broadband fluxes by diffuse continuum and line emission from the outer broad line region \citep[BLR;][]{Korista_Goad_2001, netzer2022}, which introduces additional lags and distorts the lag-spectrum, challenging the assumption that $\tau_{\lambda} \, \propto \, \lambda^{\beta}$.

To overcome these limitations, we developed a self-consistent model that incorporates intrinsic disk reprocessing and BLR-induced contamination, derived from a physically motivated, radiation-pressure–regulated BLR kinematic framework based on the Failed Radiatively Accelerated Dusty Outflow (FRADO) model \citep{czhr2011, naddaf2021} into the observed lag-spectrum modeling \citep[see][for details]{vikram2025, 2026arXiv260627195M}. Strong theoretical support for the model comes from \citet{owen2025}. This framework yields a more realistic description of AGN continuum variability and, through joint fitting of lag-spectra and spectral energy distributions (SEDs), enables estimation of the Hubble constant ($H_0$). Applying this method across a broad AGN sample provides a new approach to probing the Hubble tension.

Motivated by this potential, we launched the Hubble constant constraint through AGN Light curves Observation (HALO) program in 2024 at the Rolf Chini Cerro Murphy Observatory (in short OCM\footnote{\url{https://ocm.camk.edu.pl/}}) in Chile.

\section{Objective, Sample and Observation strategy}

\begin{figure}
  \centering
  \includegraphics[width=0.32\linewidth]{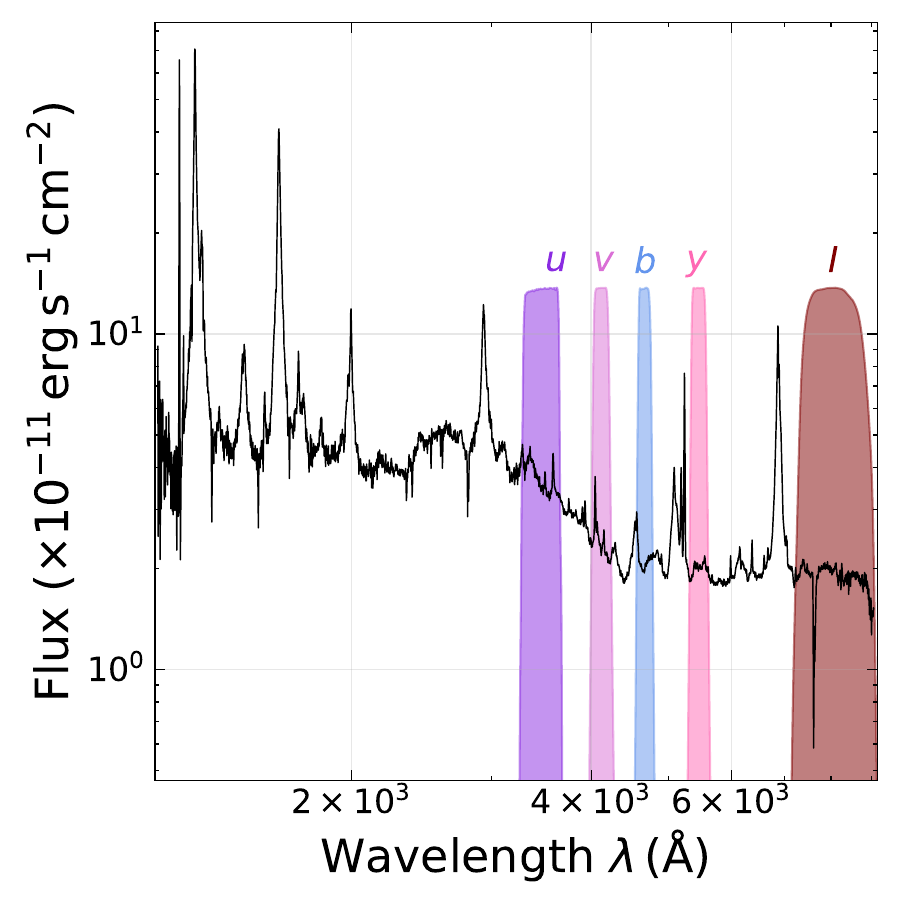}
    \includegraphics[width=0.32\linewidth]{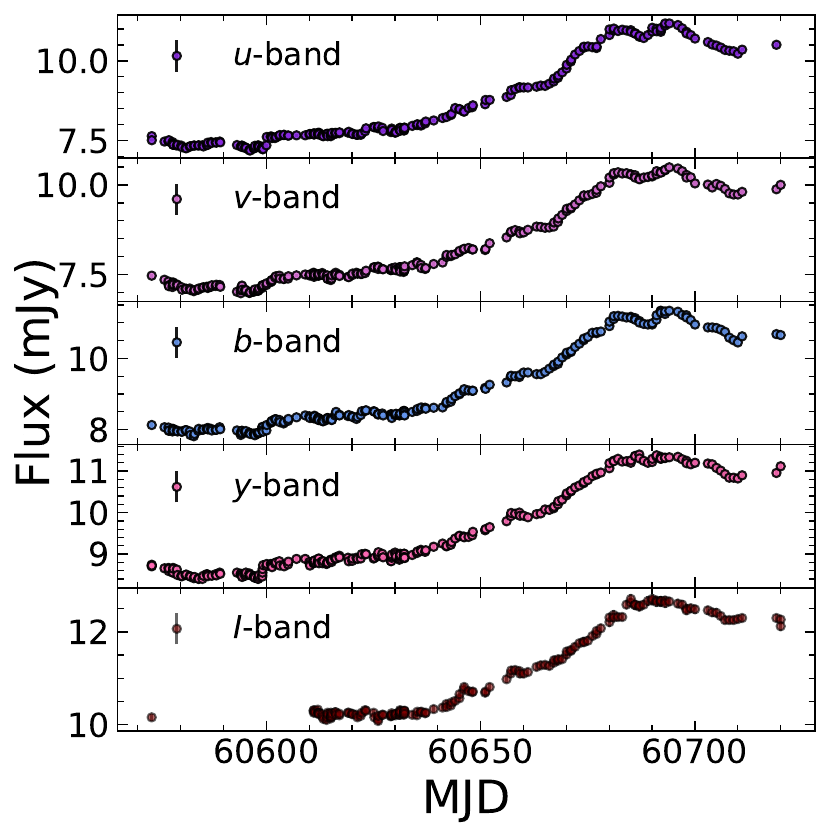}
    \includegraphics[width=0.32\linewidth]{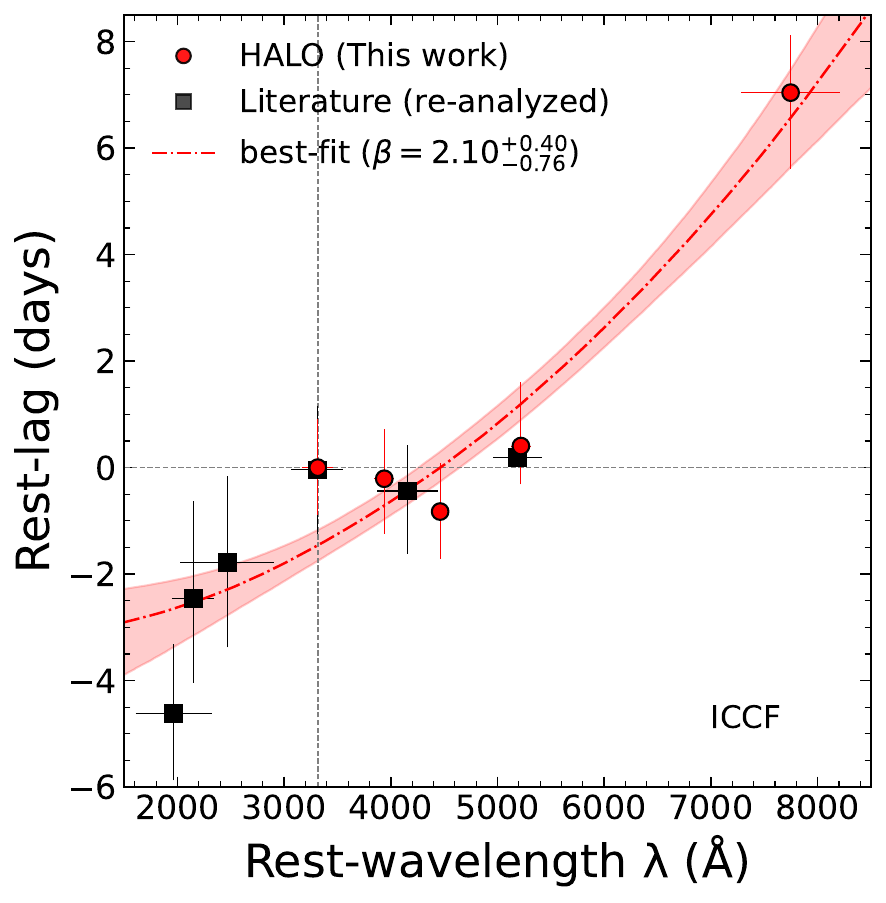}
    \caption{Exemplary HALO monitoring results for Fairall~9. Left: OCM filter transmission curves overlaid on the UV and optical spectra of Fairall~9 from HST/FOS and the ESO 1.5 m telescope. Middle: HALO light curves in the five filters. Right: Lag-spectrum showing inter-band delays as a function of wavelength, with red points from HALO and black points from our re-analysis of published data from literature. Inter-band delays were derived using the Interpolated Cross-Correlation Function method \citep[ICCF;][]{1986ApJ...305..175G, 1998PASP..110..660P}. The lag-spectrum shows excess delays in the OCM $u$ and Swift $U$ bands from BLR Balmer-continuum contamination, and in the $I$ band from torus emission, both causing significant deviations from the expected standard disk model behavior. See also \citet{2026A&A...706A.176M}.}
    \label{fig1}
\end{figure}

With its ability to obtain consecutive observations in five filters, the OCM enables high-quality continuum$-$RM measurements and construction of the lag-spectrum. By simultaneously fitting the lag-spectrum and SED with our newly developed AGN model, we can estimate $H_{0}$, the primary aim of this program.

Our parent sample contains $\sim$ 150 AGNs spanning a wide redshift range: $\sim$ 50 at $z < 0.1$, another 50 within $0.1 < z < 2.0$, and 50 between $2.0 < z < 3.0$. From this we selected a best-sample of Type~1 AGNs covering a broad luminosity range ($10^{42} \, {\rm erg \, s^{-1}} < L_{5100} < 10^{47.4} \, {\rm erg \, s^{-1} 
 }$), focusing on sources with H$\beta$ BLR size measurements from spectroscopic--RM \citep{2015PASP..127...67B,2024ApJS..275...13W}. This subset provides both black hole mass and $R_{\rm BLR}$ and includes objects with pronounced flux variability. Together with the empirical correlation between continuum-emission and BLR sizes, these data define the optimal cadence and baseline for our monitoring program. Based on visibility from OCM in Chile, we initiated observations of about 20 AGNs from this best-sample with declinations $\lesssim 0^\circ$. 

To mitigate contamination from broad emission features within the filter bandpasses while maximizing the accretion disk continuum, we selected four Strömgren intermediate-band filters ($u$, $v$, $b$, $y$) and one Johnson--Cousins $I$ filter. Observations are conducted on sub-day or daily cadence, depending on source luminosity.

\begin{table*}
\centering

 \caption{Light curve properties and results from the lag analysis for Fairall~9.}
 \label{tab:var}

\resizebox{9cm}{!}{
\fontsize{4pt}{4pt}\selectfont
\begin{tabular}{lcccr} \hline \hline

Filter & $t_{\rm span}$ & N &  $\Delta t$ & $\tau_{\rm delay}$ 
\\ 
 & (days) &  & (days) & (days)  \\
(1) & (2) & (3) & (4) & (5)
\\ \hline \\

OCM-$u$ & 146 &  236 & 0.25 & $-0.00_{-0.92}^{+0.96}$   \\
OCM-$v$ & 147 &  240 & 0.23 & $-0.22_{-1.08}^{+0.97}$  \\
OCM-$b$  & 147 &  239 & 0.24 & $-0.86_{-0.93}^{+0.95}$ \\
OCM-$y$  & 147 &  241 & 0.23 & $0.42_{-0.75}^{+1.27}$ \\
OCM-$I$  & 147 &  164 & 0.77 & $7.37_{-1.50}^{+1.14}$  \\
\hline \\
Swift-$W2$ & 1007 & 550  & 1.13 & $-4.83_{-1.32}^{+1.35}$ \\
Swift-$M2$ & 1007 & 507  & 1.20 & $-2.57_{-1.66}^{+1.91}$ \\
Swift-$W1$ & 1007 & 506  & 1.20 & $-1.87_{-1.67}^{+1.70}$ \\
Swift-$U$ & 1007 & 516  & 1.14 & $-0.05_{-1.26}^{+1.26}$ \\
Swift-$B$ & 1007 & 578  & 1.07 & $-0.46_{-1.24}^{+0.90}$ \\
Swift-$V$ & 1007 & 610  & 1.07 & $0.20_{-0.16}^{+0.16}$ \\
\hline

\end{tabular}
}

\vspace{0.1cm}
\begin{minipage}{\textwidth}
\footnotesize
\textit{Notes.} Columns are: (1) name of filter, 
(2) temporal baseline of the light curve, (3) number of data points, 
(4) median cadence, and (5) inter-band time delay ($\tau_{\rm delay}$) in the observed-frame, calculated using ICCF. The time delays for the OCM and Swift light curves are measured relative to the OCM-$u$ and Swift-$U$ bands, respectively. For further details, see \citet{2026A&A...706A.176M}.
\end{minipage}

\end{table*}

To date, we have completed photometric monitoring for three targets including Fairall~9. Figure~\ref{fig1} shows an exemplary results for Fairall~9 from HALO program. In addition, we incorporated supplementary light curve data for Fairall\,9 from Swift observations, as compiled by \citet{edelson2024}. The light curve properties and the results of the inter-band time delay analysis are summarized in Table\,\ref{tab:var}.

\section{Expected Outcomes}

Constructing a lag-spectrum for each AGN deepens our understanding of accretion-disk physics by characterizing the origin of inter-band time delays. By applying our newly developed model to fit the observed lag-spectrum, while disentangling BLR and torus contamination, we can estimate $H_0$ from each disk-mapping measurement.  Further discussion of the potential of quasar light-echo studies for cosmological applications is presented in the accompanying paper in this volume (B. Czerny et al., \textit{Quasars for Cosmology}). This, in turn, enables a statistical analysis of the recovered $H_0$ across a broad redshift range, providing insight into systematic uncertainties. A first application of our model to archival data for NGC~5548 yielded an estimate of $H_0$ = $66.9^{+10.6}_{-2.1}\,\rm km\,s^{-1}\,Mpc^{-1}$, with an uncertainty of approximately 10--15\% \citep{vikram2025}. Applying the model to our second source, Fairall~9, using combined OCM and Swift observations, we obtained $H_{0}=72.4_{-3.7}^{+3.4}\,\rm km\,s^{-1}\,Mpc^{-1}$, with a measurement precision of approximately 5\% \citep{2026arXiv260627195M}. Building on these results, continued monitoring of our targets with OCM within the HALO program will enable us to assess the consistency of $H_0$ measurements across different sources, quantify the dominant astrophysical systematics, and derive a combined constraint on $H_0$ with substantially reduced statistical uncertainty. Moreover, combining our observations with monitoring from the Legacy Survey of Space and Time (LSST), covering more than 1000 AGNs  per Deep Drilling Field \citep{2022ApJS..262...49K} could reduce the uncertainty to below 0.2\%, offering a promising avenue for addressing the Hubble tension.

\acknowledgements{ This project has received funding from the European Research Council (ERC) under the European Union’s Horizon 2020 research and innovation program (grant agreement No. [951549]). HALO observations are being carried out using the 0.6\,m Wojtek Krzemi\'nski  Telescope (WK06) at the Rolf Chini Cerro Murphy Observatory (OCM) in Chile. The OCM light curve data of Fairall~9 are available in \citet{2026A&A...706A.176M}. We used the Swift light curve data and the combined UV and optical spectrum of Fairall~9 presented by \citet{edelson2024}.}

\bibliographystyle{ptapap}
\bibliography{mandal}

\end{document}